\documentclass[pra,aps,twocolumn,showpacs]{revtex4}
\usepackage{graphicx,bbm}

\begin{document}

\title{Distinguishing between optical coherent states with
    imperfect detection}
\author{JM Geremia}
\email{jgeremia@Caltech.EDU}
\affiliation{Physics and Control \&
Dynamical Systems, California Institute of Technology, Pasadena,
CA 91125}
\date{\today}


\begin{abstract}
Several proposed techniques for distinguishing between optical
coherent states are analyzed under a physically realistic model of
photodetection.  Quantum error probabilities are derived for the
Kennedy receiver, the Dolinar receiver and the unitary rotation
scheme proposed by Sasaki and Hirota for sub-unity detector
efficiency. Monte carlo simulations are performed to assess the
effects of detector dark counts, dead time, signal processing
bandwidth and phase noise in the communication channel. The
feedback strategy employed by the Dolinar receiver is found to
achieve the Helstrom bound for sub-unity detection efficiency and
to provide robustness to these other detector imperfections making
it more attractive for laboratory implementation than previously
believed.
\end{abstract}

\maketitle

\section{Introduction} \label{Section::Introduction}

Communication is subject to quantum mechanical indeterminism even
when the transmitted information is entirely classical.  This
potentially counter-intuitive property results from the fact that
information must be conveyed through a physical medium--- a
\textit{communication channel}--- that is unavoidably governed by
quantum mechanics. From this perspective, the \textit{sender}
encodes information by preparing the channel into a well-defined
quantum state, $\hat\rho$, selected from a predetermined alphabet,
$\mathcal{A}\equiv \{\hat\rho_0, \ldots,\hat\rho_M\}$, of
codewords.  The \textit{receiver}, following any relevant signal
propagation, performs a measurement on the channel to ascertain
which state was transmitted by the sender.

A quantum mechanical complication arises when the states in
$\mathcal{A}$ are not orthogonal, as no measurement can
distinguish between overlapping quantum states without some
ambiguity \cite{vonNeumann1955,Holevo1973,Helstrom1976,Fuchs1996}.
This uncertainty in determining the channel state translates into
a non-zero probability that the receiver will misinterpret the
transmitted codeword and produce a communication error. While it
would seem obvious that the sender should simply adopt an alphabet
of orthogonal states, it is rarely practicable to communicate
under such ideal conditions \cite{Shapiro1978,Hirota1995}. Even
when it is possible for the sender to transmit orthogonal
codewords, inevitable imperfections in the channel including
decoherence and energy dissipation quickly damage that
orthogonality. In some cases, the classical information capacity
of a noisy channel is actually maximized by a nonorthogonal
alphabet \cite{Fuchs1997}.

When developing a communication system to operate at the highest
feasible rate given fixed channel properties and a constrained
capability for state preparation, the objective is to minimize the
communication error by designing a ``good'' receiver.
Distinguishing between nonorthogonal states is a pervasive problem
in quantum information theory \cite{Peres1991,FuchsThesis}
addressed mathematically by optimizing a state-determining
measurement over all positive operator valued measures (POVMs)
\cite{Davies1970,Helstrom1976,Kraus1983}. This general approach
can be applied to communication; however, arbitrary POVMs are
rarely straightforward to implement in the laboratory. Therefore,
a ``good'' receiver must balance quantum mechanical optimality
with implementability and robust performance under realistic
experimental conditions.

For example, the optical field produced by a laser provides a
convenient quantum system for carrying information. Of course,
optical coherent states are not orthogonal and cannot be
distinguished perfectly by photodetection.  While the overlap
between different coherent states can be reduced by employing
large amplitudes, power limitations often restrict $\mathcal{A}$
to the small-amplitude regime where quantum effects dominate. This
is especially true in situations (such as optical fibers) where
the communication medium behaves nonlinearly at high power, as
well as for long distance communication where signals are
substantially attenuated, including deep space transmission.

Motivated by these experimental considerations, optimizing a
communication process based on small-amplitude optical coherent
states and photodetection has been an active subject since the
advent of the laser
\cite{Kennedy1972,Dolinar1973,Hirota1995,Belavkin1995}. Kennedy
initially proposed a receiver based on simple photon counting to
distinguish between two different coherent states
\cite{Kennedy1972}. However, the Kennedy receiver error
probability lies above the quantum mechanical minimum
\cite{Helstrom1976} (or Helstrom bound) and this prompted Dolinar
to devise a measurement scheme capable of achieving the quantum
limit \cite{Dolinar1973}. Dolinar's receiver, while still based on
photon counting, approximates an optimal POVM by adding a local
feedback signal to the channel; but, this procedure has often been
deemed impractical \cite{Hirota1996a} due to the need for
real-time adjustment of the local signal following each photon
arrival. As a result, Sasaki and Hirota later proposed an
alternative receiver that applies an open-loop unitary
transformation to the incoming coherent state signals to render
them more distinguishable by simple photon counting
\cite{Hirota1995,Hirota1996a,Hirota1996b} .

However, recent experimental advances in real-time quantum-limited
feedback control \cite{Armen2002,Geremia2004a,Geremia2004b}
suggest that the Dolinar receiver may be more experimentally
practical than previously believed. The opinion that feedback
should be avoided in designing an optical receiver is grounded in
the now-antiquated premise that real-time adaptive quantum
measurements are technologically inaccessible. Most arguments in
favor of passive devices have been based on idealized receiver
models that assume, for example, perfect photon counting
efficiency.  A fair comparison between open and closed-loop
receivers should take detection error into account--- feedback
generally increases the robustness of the measurement device in
exchange for the added complexity.

Here, we consider the relative performance of the Kennedy, Dolinar
and Sasaki-Hirota receivers under \textit{realistic} experimental
conditions that include: (1) sub-unity quantum efficiency, where
it is possible for the detector miscount incoming photons, (2)
non-zero dark-counts, where the detector can register photons even
in the absence of a signal, (3) non-zero dead-time, or finite
detector recovery time after registering a photon arrival, (4)
finite bandwidth of any signal processing necessary to implement
the detector, and (5) fluctuations in the phase of the incoming
optical signal.

\section{Binary Coherent State Communication}
    \label{Section::CoherentStateCommunication}

An optical binary communication protocol can be implemented via
the alphabet consisting of two pure coherent states,
$\hat\rho_0=|\Psi_0\rangle\langle\Psi_0|$, and
$\hat\rho_1=|\Psi_1\rangle\langle\Psi_1|$.  Without loss of
generality, we will assume that logical 0 is represented by the
vacuum,
\begin{equation} \label{Equation::Psi0}
    \Psi_0(t) = 0 \,,
\end{equation}
and that logical 1 is represented by
\begin{equation} \label{Equation::Psi1}
    \Psi_1(t) = \psi_1(t) \exp\left[-i (\omega t + \varphi)
    \right] + \mathrm{c.c.},
\end{equation}
where $\omega$ is the frequency of the optical carrier and
$\varphi$ is (ideally) a fixed phase.  The envelope function,
$\psi_1(t)$, is normalized such that
\begin{equation}
   \int_0^{T} |\psi_1(t)|^2 \, dt =  \bar{N} \,,
\end{equation}
where $\bar{N}$ is the mean number of photons to arrive at the
receiver during the measurement interval, $0\le t \le T$.  That
is, $\hbar \omega |\psi_1(t)|^2$ is the instantaneous average
power of the optical signal for logical 1.

This alphabet, $\mathcal{A}=\{\hat\rho_0,\hat\rho_1 \}$, is
applicable to both amplitude and phase-shift keyed communication
protocols as it is always possible to transform between the two by
combining the incoming signal with an appropriate local
oscillator. That is, amplitude keying with
$\mathcal{A}=\{|0\rangle,|\alpha\rangle\}$ (for some coherent
state $|\alpha\rangle$ with amplitude, $\alpha$) is equivalent to
the phase-shift keyed alphabet,
$\{|-\frac{1}{2}\alpha\rangle,|\frac{1}{2}\alpha\rangle\}$, via a
displacement, $\hat D\left[-\frac{1}{2}\alpha\right] \equiv \exp
(-\frac{1}{2}(\alpha \hat a^\dagger - \alpha^* \hat a))$, where
$\hat{a}^\dagger$ and $\hat{a}$ are the creation and annihilation
operators for the channel mode.  Similarly, if
$|\Psi_0\rangle\ne|0\rangle$, a simple displacement can be used
restore $|\Psi_0\rangle$ to the vacuum state.

\subsection{The Quantum Error Probaility}

The coherent states, $\hat\rho_0$ and $\hat\rho_1$, are not
orthogonal, so it is impossible for a receiver to identify the
transmitted state without sometimes making a mistake. That is, the
receiver attempts to ascertain which state was transmitted by
performing a quantum measurement, $\Upsilon$, on the channel.
$\Upsilon$ is described by an appropriate POVM represented by a
complete set of positive operators \cite{Peres1990},
\begin{equation}
    \sum_n \hat\Upsilon_n = \hat\mathbbm{1}\, \quad
    \hat\Upsilon_n \ge 0\, ,
\end{equation}
where $n$ indexes the possible measurement outcomes. For binary
communication, it is always possible (and optimal) for the
receiver to implement the measurement as a decision between two
hypotheses: $(H_0)$, that the transmitted state is $\hat\rho_0$,
selected when the measurement outcome corresponds to
$\hat\Upsilon_0$, and $(H_1)$, that the transmitted state is
$\hat\rho_1$, selected when the measurement outcome corresponds to
$\hat\Upsilon_1$.

Given the positive operators, $\hat\Upsilon_0 + \hat\Upsilon_1 =
\hat\mathbbm{1}$, there is some chance that the receiver will
select the null hypothesis, $H_0$, when $\hat\rho_1$ is actually
present,
\begin{equation}
   p(H_0 | \hat\rho_1) = \mathrm{tr}[\hat\Upsilon_0 \hat\rho_1]
   = \mathrm{tr}[(\hat\mathbbm{1}-\hat\Upsilon_1) \hat\rho_1 ]\,.
\end{equation}
And, it will sometimes select $H_1$ when $\hat\rho_0$ is present,
\begin{equation}
   p(H_1 | \hat\rho_0) =
       \mathrm{tr}[\hat\Upsilon_1 \hat\rho_0] \,.
\end{equation}
The total receiver error probability depends upon the choice of
$\hat\Upsilon_0$ and $\hat\Upsilon_1$ and is given by
\begin{equation}
    p[\hat\Upsilon_0,\hat\Upsilon_1] =
    \xi_0 p(H_1|\hat\rho_0) + \xi_1 p(H_0|\hat\rho_1) \,.
\end{equation}
Here, $\xi_0=p_0(\hat\rho_0)$ and $\xi_1 = p_0(\hat\rho_1)$ are
the probabilities that the sender will transmit $\hat\rho_0$ and
$\hat\rho_1$ respectively; they reflect the prior knowledge that
enters into the hypothesis testing process implemented by the
receiver, and in many cases $\xi_0= \xi_1 = 1/2$.

Minimizing the receiver measurement over POVMs (over
$\hat\Upsilon_0$ and $\hat\Upsilon_1$) leads to a quantity known
as the \textit{quantum error probability},
\begin{equation}
  P_\mathrm{H} \equiv \min_{\hat\Upsilon_0,\hat\Upsilon_1}
    p[\hat\Upsilon_0,\hat\Upsilon_1] \,,
\end{equation}
also referred to as the Helstrom bound.  $P_\mathrm{H}$ is the
smallest physically allowable error probability, given the overlap
between $\hat\rho_0$ and $\hat\rho_1$.

\subsubsection{The Helstrom Bound}

Helstrom demonstrated that minimizing the receiver error
probability,
\begin{eqnarray}
  p[\hat\Upsilon_0, \hat\Upsilon_1] & = &
    \xi_0 \mathrm{tr}\left[\hat\Upsilon_1 \hat\rho_0 \right] +
    \xi_1 \mathrm{tr}\left[(\hat\mathbbm{1}-\hat\Upsilon_1) \hat\rho_1
    \right] \\
    & = & \xi_1 + \mathrm{tr}\left[ \hat\Upsilon_1 (\xi_0 \hat\rho_0
    - \xi_1 \hat\rho_1 ) \right]
\end{eqnarray}
is accomplished by optimizing
\begin{equation}
  \min_{\hat\Upsilon_1} \mathrm{tr}[\hat\Upsilon_1 \hat\Gamma]\,, \quad \hat\Gamma = \xi_0 \hat\rho_0 - \xi_1 \hat\rho_1
\end{equation}
over $\hat\Upsilon_1$ subject to $0 \le \hat\Upsilon_1 \le
\hat\mathbbm{1}$ \cite{Helstrom1976}.  Given the spectral
decomposition,
\begin{equation}
    \hat\Gamma = \sum_n \lambda_n |n \rangle\langle n|\,
\end{equation}
where the $\lambda_n$ are the eigenvalues of $\hat\Gamma$, the
resulting Helstrom bound can be expressed as \cite{Helstrom1967}
\begin{equation}
  P_\mathrm{H} = \xi_1 + \sum_{\lambda_n < 0} \lambda_n \,.
\end{equation}
For pure states, where $\hat\rho_0=|\Psi_0\rangle\langle\Psi_0|$
and $\hat\rho_1=|\Psi_1\rangle\langle\Psi_1|$, $\hat\Gamma$ has
two eigenvalues of which only one is negative,
\begin{equation}
\lambda_- =  \frac{1}{2}\left( 1 - \sqrt{1
   - 4 \xi_0 \xi_1 |\langle \Psi_0 | \Psi_1 \rangle|^2} \right) -
   \xi_1
   > 0 \,,
\end{equation}
and the quantum error probability is therefore \cite{Helstrom1976}
\begin{equation} \label{Equation::PerfectHelstrom}
  P_\mathrm{H} = \frac{1}{2} \left(1-\sqrt{1-4 \xi_0 \xi_1 | \langle \Psi_1 |
  \Psi_0 \rangle |^2} \right) \,.
\end{equation}

The Helstrom bound is readily evaluated for coherent states by
employing the relation \cite{Glauber1963},
\begin{equation}
   |\alpha\rangle = e^{-|\alpha|^2/2} \sum_{n=0}^\infty
    \frac{\alpha^n}{\sqrt{n!}} |n\rangle \, ,
\end{equation}
to compute the overlap between $|\Psi_1\rangle$ and
$|\Psi_0\rangle$,
\begin{equation} \label{Equation::CoherentOverlap}
    c_0 \equiv \langle \Psi_1 | \Psi_0 \rangle =
    e^{-\bar{N}/2} \, .
\end{equation}
It is further possible to evaluate the Helstrom bound for
imperfect detection.  Coherent states have the convenient property
that sub-unity quantum efficiency is equivalent to an ideal
detector masked by a beam-splitter with transmission coefficient,
$\eta \le 1$, to give
\begin{equation} \label{Equation::RealHelstrom}
  P_\mathrm{H}(\eta) =
  \frac{1}{2} \left(1-\sqrt{1-4 \xi_0 \xi_1
    c_0^{2 \eta}} \right) \,.
\end{equation}
This result and Eq.\ (\ref{Equation::PerfectHelstrom}) indicate
that there is a finite quantum error probability for all choices
of $|\Psi_1\rangle$, even when an optimal measurement is
performed.

\subsection{The Kennedy Receiver}

Kennedy proposed a near-optimal receiver that simply counts the
number of photon arrivals registered by the detector between $t=0$
and $T$. It decides in favor of $H_0$ when the number of clicks is
zero, otherwise $H_1$ is chosen. This hypothesis testing procedure
corresponds to the measurement operators,
\begin{eqnarray}
    \hat\Upsilon_0 & = & |0\rangle\langle0| \\
    \hat\Upsilon_1 & = & \sum_{n=1}^\infty |n\rangle\langle n| \,
\end{eqnarray}
where $|n\rangle$ are the eigenvectors of the number operator,
$\hat{N}=\hat{a}^\dagger \hat{a}$.

The Kennedy receiver has the property that it always correctly
selects $H_0$ when the channel is in $\hat\rho_0$, since the
photon counter will never register photons when the vacuum state
is present (ignoring background light and detector dark-counts for
now). Therefore, $p(H_1|\hat\rho_0)=0$, however,
\begin{equation}
  p(H_0 | \hat\rho_1) \equiv \mathrm{tr}[\hat\Upsilon_0 \hat\rho_1]
    = |\langle 0 | \Psi_1
\rangle|^2
\end{equation}
is non-zero due to the finite overlap of all coherent states with
the vacuum.  The Poisson statistics of coherent state photon
numbers allows for the possibility that zero photons will be
recorded even when $\hat\rho_1$ is present.

Furthermore, an imperfect detector can misdiagnose $\hat\rho_1$ if
it fails to generate clicks for photons that do arrive at the
detector. The probability for successfully choosing $H_1$ when
$\hat\rho_1$ is present is given by,
\begin{equation}
    p_\eta(H_1|\hat\rho_1) = \sum_{n=1}^\infty \sum_{k=1}^\infty
       p(n,k) |\langle n | \alpha\rangle|^2
\end{equation}
where the Bernoulli distribution,
\begin{equation}
  p(n,k) = \frac{n!}{k!(n-k)!} \eta^{k} (1-\eta)^{n-k}
\end{equation}
gives the probability that a detector with quantum efficiency,
$\eta$, will register $k$ clicks when the actual number of photons
is $n$. The resulting Kennedy receiver error,
\begin{equation} \label{Equation::KennedyError}
    P_\mathrm{K}(\eta)  =  1 - p_\eta(H_1|\hat\rho_1)
                  =  \xi_1 c_0^{2\eta} \, .
\end{equation}
asymptotically approaches the Helstrom bound for large signal
amplitudes, but is larger for small photon numbers.

\subsection{The Sasaki-Hirota Receiver}

Sasaki and Hirota proposed that it would be possible to achieve
the Helstrom bound using simple photon counting by applying a
unitary transformation to the incoming signal states prior to
detection \cite{Hirota1995,Hirota1996a,Hirota1996b}.  They
considered rotations,
\begin{equation} \label{Equation::Unitary}
  \hat{U}[\theta] = \exp\left[\theta
    ( |\Psi_0^\prime\rangle\langle \Psi_1^\prime| -
    |\Psi_1^\prime\rangle\langle\Psi_0^\prime| ) \right]\,,
\end{equation}F
generated by the transformed alphabet, $\mathcal{A}^\prime$,
\begin{equation}
  |\Psi_0^\prime\rangle = | \Psi_0 \rangle, \quad |\Psi_1^\prime\rangle =
  \frac{|\Psi_1\rangle - c_0 |\Psi_0\rangle}{\sqrt{1-c_0^2}} \,,
\end{equation}
obtained from Gram-Schmidt orthogalization of $\mathcal{A}$.  The
rotation angle, $\theta \in \mathbbm{R}$, is a parameter that must
be optimized in order to achieve the Helstrom bound.

Application of $\hat{U}[\theta]$ on the incoming signal states
(which belong to the original alphabet, $\mathcal{A}$) leads to
the transformed states,
\begin{eqnarray}
  \hat{U}[\theta]|\Psi_0\rangle & = &
    \left(\cos\theta + \frac{c_0\sin\gamma}{\sqrt{1-c_0^2}} \right)
    | \Psi_0 \rangle  \nonumber \\
    & & - \frac{\sin\theta}{\sqrt{1-c_0^2}} | \Psi_1
    \rangle
\end{eqnarray}
and
\begin{eqnarray}
  \hat{U}[\theta]|\Psi_1\rangle & = &
  \frac{\sin\theta}{\sqrt{1-c_0^2}} |\Psi_0\rangle \\ & & +
  \frac{\cos\theta\sqrt{1-c_0^2}-c_0\sin\theta}{\sqrt{1-c_0^2}} |
  \Psi_1\rangle \nonumber \,.
\end{eqnarray}

Since $|\Psi_0^\prime\rangle$ is the vacuum state, hypothesis
testing can still be performed by simple photon counting. However,
unlike the Kennedy receiver, it is possible to misdiagnose
$\hat\rho_0$ since $\hat{U}[\theta]|\Psi_0\rangle$ contains a
non-zero contribution from $|\Psi_1\rangle$.  The probability for
a false-positive detection by a photon counter with efficiency,
$\eta$, is given by
\begin{eqnarray}
  p_\eta^\theta(H_1|\hat\rho_0) & = & \sum_{n=1}^\infty \sum_{k=1}^\infty
    p(n,k) |\langle n | \hat{U}[\theta] | \Psi_0\rangle|^2 \\
    & = & \frac{c_0^{2\eta}-1}{c_0^2-1} \sin^2\theta
\end{eqnarray}
which is evaluated by recognizing that
\begin{eqnarray}
  \langle n | \hat{U}[\theta] | \Psi_0 \rangle & = &
    \left[ \cos\theta + \frac{c_0 \sin\theta}{\sqrt{1-c_0^2}}
    \right] \delta_{n,0} \nonumber \\ &&
    - \frac{c_0 \alpha^n \sin\theta}{\sqrt{n!(1-c_0^2)}} \,,
\end{eqnarray}
where $\alpha$ is the (complex) amplitude of $|\Psi_1\rangle$. The
probability for correct detection can be similarly obtained to
give
\begin{eqnarray}
  p_\eta^\theta(H_1|\hat\rho_1) & = & \sum_{n=1}^\infty \sum_{k=1}^\infty
    p(n,k) | \langle n | \hat{U}[\theta] | \Psi_1 \rangle \\
    & = & \frac{c_0^{2\eta}-1}{c_0^2-1}\left[c_0 \sin\theta -
    \sqrt{1-c_0^2}\cos\theta  \right]^2
\end{eqnarray}
by employing the relationship,
\begin{eqnarray}
  \langle n | \hat{U}[\theta] | \Psi_1 \rangle & = &
  \left[ c_0 \cos\theta - \frac{c_0^2\alpha^n\sin\theta}{
  \sqrt{n!(1-c_0^2)}} \right] \nonumber \\ & &+ \frac{\sin\theta}{\sqrt{1-c_0^2}}
  \delta_{n,0} \,.
\end{eqnarray}

The total Sasaki-Hirota receiver error is given by the weighted
sum,
\begin{equation} \label{Equation::SHError}
  P_\mathrm{SH}(\eta,\theta) = \xi_0 p_\eta^\theta(H_1|\hat\rho_0) + \xi_1 \left[
  1-p_\eta^\theta(H_1|\hat\rho_1) \right]
\end{equation}
and can be minimized over $\theta\in\mathbbm{R}$ to give
\begin{equation}
  \theta = - \tan^{-1} \sqrt{\frac{\sqrt{1-4 \xi_0 \xi_1 c_0^2} - 1 + 2
  \xi_1 c_0^2}{\sqrt{1-4 \xi_0 \xi_1 c_0^2} + 1 - 2 \xi_1 c_0^2}}
  \,.
\end{equation}
For perfect detection efficiency, $\eta=1$, Eq.\
(\ref{Equation::SHError}) is equivalent to the Helstrom bound;
however, for $\eta<1$, it is larger.

\subsection{The Dolinar Receiver}

The Dolinar receiver takes a different approach to achieving the
Helstrom bound with a photon counting detector; it utilizes an
adaptive strategy to implement a feedback approximation to the
Helstrom POVM \cite{Dolinar1973,DolinarThesis}.  Dolinar's
receiver operates by combining the incoming signal, $\Psi(t)$,
with a separate local signal,
\begin{equation}
  U(t) = u(t) \exp\left[-i (\omega t + \phi) \right] + \mathrm{c.c}
  \,,
\end{equation}
such that the detector counts photons with total instantaneous
mean rate,
\begin{equation}
    \Phi(t) = |\psi(t)+u(t)|^2  \,.
\end{equation}
Here, $\psi(t)=0$ when the channel is in the state $\hat\rho_0$,
and $\psi(t)=\psi_1(t)$ when the channel is in $\hat\rho_1$ [refer
to Eqs.\ (\ref{Equation::Psi0}) and (\ref{Equation::Psi1})].

The receiver decides between hypotheses $H_0$ and $H_1$ by
selecting the one that is more consistent with the record of
photon arrival times observed by the detector given the choice of
$u(t)$.  $H_1$ is selected when the ratio of conditional arrival
time probabilities,
\begin{equation}
    \Lambda = \frac{p_\eta\left[\hat\rho_1 | t_1,\ldots,t_n, u(t) \right]}
    {p_\eta\left[\hat\rho_0 | t_1, \ldots, t_n,u(t)\right]}\, ,
\end{equation}
is greater than one; otherwise it is assumed that $\hat\rho_0$ was
transmitted.  The conditional probabilities,
$p_\eta\left[\hat\rho_i|t_1,\ldots,t_n,u(t)\right]$, reflect the
likelihood that $n$ photon arrivals occur precisely at the times,
$\{t_1,\ldots,t_n\}$, given that: the channel is in the state,
$\hat\rho_i$, the feedback amplitude is $u(t)$, and the detector
quantum efficiency is $\eta$.

We see that this decision criterion based on $\Lambda$ is
immediately related to the error probabilities,
\begin{equation} \label{Equation::LambdaH1}
    \Lambda =
    \frac{p_\eta\left[H_1|\hat\rho_1,u(t)\right]}{p_\eta\left[H_1\hat\rho_0,u(t)\right]} =
    \frac{1-p_\eta\left[H_0|\hat\rho_1,u(t)\right]}{p_\eta\left[H_1|\hat\rho_0,u(t)\right]},
\end{equation}
when $\Lambda>1$ (i.e., the receiver definitely selects $H_1$),
and
\begin{equation} \label{Equation::LambdaH0}
   \Lambda =
    \frac{p_\eta\left[H_0|\hat\rho_1,u(t)\right]}{p_\eta\left[H_0|\hat\rho_0,u(t)\right]}=
    \frac{p_\eta\left[H_0|\hat\rho_1,u(t)\right]}{1-p_\eta\left[H_1|\hat\rho_0,u(t)\right]}
\end{equation}
when $\Lambda < 1$ (i.e., the receiver definitely selects $H_0$).
Similarly, the likelihood ratio, $\Lambda$, can be re\"{e}xpressed
in terms of the photon counting distributions frequently
encountered in quantum optics by employing Bayes' rule,
\begin{eqnarray} \label{Equation::LambdaUseful}
    \Lambda & = & \frac{p_\eta\left[t_1,\ldots,t_n|\hat\rho_1,u(t)\right] p_0(\hat\rho_1) }
    { p_\eta\left[t_1,\ldots,t_n | \hat\rho_0, u(t)\right] p_0(\hat\rho_0) } \\
    & = & \frac{\xi_1}{\xi_0} \frac{ p_\eta\left[t_1,\ldots,t_n| \hat\rho_1,u(t)\right]}
    { p_\eta\left[t_1,\ldots,t_n|\hat\rho_0,u(t)\right]} \,,
\end{eqnarray}
where the $p_\eta\left[t_1,\ldots,t_n|\hat\rho_i,u(t)\right]$ are
the exclusive counting densities,
\begin{equation} \label{Equation::ArrivalTimeProbability}
  p_\eta\left[t_1,\ldots,t_n|\hat\rho_i,u(t)\right] =
    \prod_{k=1}^{n+1} w_\eta\left[t_k | \hat\rho_i,u(t)\right]\,.
\end{equation}
Here, $t_0=0$, $t_{n+1}=T$, and $w_\eta\left[t_k |
\hat\rho_i,u(t)\right]$ is the exponential waiting time
distribution,
\begin{equation} \label{Equation::WaitingTimeDistribution}
    w\left[t_k|\hat\rho_i,u(t)\right] = \eta \Phi(t_k) \, \exp\left(- \eta \int_{t_{k-1}}^k
    \Phi(t^\prime) \, dt^\prime\right),
\end{equation}
for optical coherent states, or the probability that a photon will
arrive at time $t_k$ and that it will be the only click during the
half-closed interval, $(t_{k-1},t_k]$ \cite{Glauber1963}.

\subsubsection{Optimal Control Problem}

The Dolinar receiver error probability,
\begin{equation}
  P_\mathrm{D}[u(t)] = \xi_0 p_\eta\left[H_1|\hat\rho_0,u(t)\right]+
    \xi_1 p_\eta\left[H_0|\hat\rho_1,u(t)\right] \,,
\end{equation}
depends upon the amplitude of the locally applied feedback field,
so the objective is to minimize $P_\mathrm{D}$ over $u(t)$. This
optimization can be accomplished \cite{DolinarThesis} via the
technique of dynamic programming \cite{Bertsekas2000}, where we
adopt an effective state-space picture given by the conditional
error probabilities,
\begin{equation}
    \mathbf{p}(t) = \left( \begin{array}{c}
        p_\eta\left[H_1|\hat\rho_0, u(t)\right](t) \\
        p_\eta\left[H_0|\hat\rho_1, u(t)\right](t) \end{array}
        \right)
\end{equation}
and define the control cost as
\begin{equation}
    \mathcal{J}[u(t)] \equiv P_\mathrm{D}[u(t)] = \xi^\mathrm{T}
    \mathbf{p} \,.
\end{equation}

The optimal control policy, $u^*(t)$, is identified by solving the
Hamilton-Jacobi-Bellman equation,
\begin{equation} \label{Equation::HJB}
    \min_{u(t)}\left[ \frac{\partial}{\partial t} \mathcal{J}[u(t)] +
     \nabla_\mathbf{p} \mathcal{J}[u(t)]^\mathrm{T} \frac{\partial}{\partial t}
     \mathbf{p}(t)  \right] = 0\, ,
\end{equation}
which is a partial differential equation for $\mathcal{J}$ based
on the requirement that $\mathbf{p}(t)$ and $u(t)$ are smooth
(continuous and differentiable) throughout the entire receiver
operation. However, like all quantum point processes, our
conditional knowledge of the system state evolves smoothly only
\textit{between} photon arrivals.

When a click is recorded by the detector, the system
probabilities, $\mathbf{p}$, can jump in a non-smooth manner.
Therefore, the photon arrival times divide the measurement
interval, $0 \le t \le T$, into segments that are only piecewise
continuous and differentiable. Fortunately, the dynamic
programming \textit{optimality principle} \cite{Bertsekas2000}
allows us to optimize $u(t)$ in a piecewise manner that begins by
minimizing $\mathcal{J}[u(t)]$ on the final segment, $[t_n,T]$. Of
course, the system state at the beginning of this segment,
$\mathbf{p}(t_n)$, depends upon the detection history at earlier
times and therefore the choice of $u(t)$ in earlier intervals. As
such, the Hamilton-Jacobi-Bellman optimization for the final
segment must hold for all possible starting states,
$\mathbf{p}(t_n) \in \mathbbm{R}^2_{[0,1]}$.  Once this is
accomplished, $u(t)$ can be optimized on the preceding segment
$[t_{n-1},t_n)$ with the assurance that any final state for that
segment will be optimally controlled on the next interval
$[t_n,T]$. This procedure is iterated in reverse order for all of
the measurement segments until the first interval, $t=[0,t_1)$,
where the initial value, $\mathbf{p}(0)=\left(
\begin{array}{cc} 1 & 0 \end{array} \right)^\mathrm{T}$, can
be unambiguously specified.

Solving the Hamilton-Jacobi-Bellman equation in each smooth
segment between photon arrivals requires the time derivatives,
$\dot\mathbf{p}(t)$, which assume a different form when $\Lambda
>1$ versus when $\Lambda< 1$. Using Eqs.\
(\ref{Equation::LambdaH1}) -- (\ref{Equation::LambdaH0}), the
coherent state waiting time distribution, and
\begin{equation}
    \Phi(t) \equiv \left( \begin{array}{c} \Phi_0(t) \\ \Phi_1(t) \end{array} \right)
    = \left( \begin{array}{c} u(t) \\ u(t) + \psi_1(t) \end{array}
    \right),
\end{equation}
we see that the smooth evolution of $\mathbf{p}(t)$ between photon
arrivals is given by
\begin{eqnarray}
   \dot{p}_0(t) & = &
        \eta p_0(t)
        \left[ \frac{d}{dt}\ln \Phi_0(t) - \Phi_0(t)\right] \\
    \dot{p}_1(t) & = &
        \eta p_1(t)
        \left[ \Phi_1(t) - \frac{d}{dt}\ln \Phi_1(t) \right] \nonumber
\end{eqnarray}
when $\Lambda>1$ and
\begin{eqnarray}
    \dot{p}_0(t) & = &
        \eta p_0(t)
        \left[ \Phi_0(t)- \frac{d}{dt}\ln \Phi_0(t) \right] \\
    \dot{p}_1(t) & = &
        \eta p_1(t)
        \left[ \frac{d}{dt}\ln \Phi_1(t) - \Phi_1(t) \right] \nonumber
\end{eqnarray}
when $\Lambda<1$.

Performing the piecewise minimization in Eq.\
(\ref{Equation::HJB}) over each measurement segment with initial
states provided by the iterative point-process probabilities in
Eq.\ (\ref{Equation::ArrivalTimeProbability}) and combining the
intervals (this is straightforward but eraser-demanding) leads to
the control policy,
\begin{equation}
    u^*_1(t)  =  -\psi_1(t)\left( 1 +
        \frac{\mathcal{J}[u_1^*(t)]}{1-2 \mathcal{J}[u_1^*(t)]} \right)
\end{equation}
for $\Lambda > 1$, where $p_\eta[H_0|\hat\rho_1,u_1^*(t)]=0$ and
\begin{eqnarray}
    \mathcal{J}[u^*_1(t)] & = &
        \xi_1 p_\eta[H_1|\hat\rho_0,u_1^*(t)] \\
        & = & \frac{1}{2}
            \left( 1 - \sqrt{1-4 \xi_0 \xi_1 e^{-\eta \bar{n}(t)}}
            \right) \,. \nonumber
\end{eqnarray}
Here, $\bar{n}(t)$ represents the average number of photons
expected to arrive at the detector by time, $t$, when the channel
is in the state, $\hat\rho_1$,
\begin{equation}
  \bar{n}(t) = \int_0^t |\psi_1(t^\prime)|^2 \, dt^\prime \,.
\end{equation}
Conversely, the optimal control takes the form,
\begin{equation}
    u^*_0(t)  =  \psi_1(t)\left(
        \frac{\mathcal{J}[u_0^*(t)]}{1-2 \mathcal{J}[u_0^*(t)]} \right)
\end{equation}
for $\Lambda < 1$, where $p_\eta[H_1|\hat\rho_0,u_0^*(t)]=0$ and
\begin{eqnarray}
    \mathcal{J}[u_0^*(t)] & = &
        \xi_1 p_\eta[H_0|\hat\rho_1,u_0^*(t)] \\
        & = & \frac{1}{2}
            \left( 1 - \sqrt{1-4 \xi_0 \xi_1 e^{-\eta \bar{n}(t)}}
            \right) \,. \nonumber
\end{eqnarray}

\subsubsection{Dolinar Hypothesis Testing Procedure}

The Hamilton-Jacobi-Bellman solution leads to a conceptually
simple procedure for estimating the state of the channel.  The
receiver begins at $t=0$ by favoring the hypothesis that is more
likely based on the prior probabilities, $p_0(0)= \xi_0$ and
$p_1(0) = \xi_1$ \footnote{If $\xi_0=\xi_1$, then neither
hypothesis is \textit{a priori} favored and the Dolinar receiver
is singular with $P_\mathrm{D}=\frac{1}{2}$.}. Assuming that
$\xi_1 \ge \xi_0$ (for $\xi_0 > \xi_0$, the opposite reasoning
applies), the Dolinar receiver always selects $H_1$ during the
initial measurement segment. The probability of deciding on $H_0$
is exactly zero prior to the first photon arrival such that an
error only occurs when the channel is actually in $\hat\rho_0$.

To see what happens when a photon does arrive at the detector, it
is necessary to investigate the behavior of $\Lambda(t)$ at the
boundary between two measurement segments. Substituting the
optimal control policy, $u^*(t)$, which alternates between
$u_1^*(t)$ and $u_0^*(t)$, into the photon counting distribution
leads to
\begin{eqnarray}
    p(t_1,\ldots,t_n|\hat\rho_i) & = & \eta^n
    \prod_{k=0}^{n+1}
    \Phi_i[u_{k|2}(0,t_1]] \times  \\
    & & \mathrm{exp} \left( -\eta \int_0^{t_1}
    \Phi_i[u_1(t^\prime_{k-1},t^\prime_k]] \,dt^\prime  - \right. \nonumber \\
    & & \left.
    \cdots  -\eta \int_{t_n}^T \Phi_i[u_{n|2}(t^\prime_{n},T^\prime]] \, dt^\prime
     \right) \nonumber \,.
\end{eqnarray}
This expression can be used to show that the limit of $\Lambda(t)$
approaching a photon arrival time, $t_k$, from the left is the
reciprocal of the limit approaching from the right,
\begin{equation}
  \lim_{t\rightarrow t_k^-} \Lambda(t) = \left[
  \lim_{t\rightarrow t_k^+} \Lambda(t) \right]^{-1} \,.
\end{equation}
That is, if $\Lambda>1$ such that $H_1$ is favored during the
measurement interval ending at $t_k$, the receiver immediately
swaps its decision to favor $H_0$ when the photon arrives.
Evidentally, the optimal control policy, $u^*(t)$, engineers the
feedback such that the photon counter is least likely to observe
additional clicks if it is correct based on its best knowledge of
the channel state at that time.  Each photon arrival invalidates
the current hypothesis and the receiver completely reverses its
decision on every click. This result implies that $H_1$ is
selected when the number of photons, $n$, is even (or zero) and
$H_0$ when the number of photons is odd.

Despite the discontinuities in the conditional probabilities,
$p_\eta[H_1|\hat\rho_0,u^*(t)]$ and
$p_\eta[H_0|\hat\rho_1,u^*(t)]$, at the measurement segment
boundaries, the total Dolinar receiver error probability,
\begin{equation}
  P_\mathrm{D}(\eta,t) = \frac{1}{2}\left(1-\sqrt{1-\xi_0\xi_1 e^{-\eta
  \bar{n}(t)}}\right),
\end{equation}
evolves smoothly since
\begin{equation}
    \lim_{t \rightarrow t_k^-} \mathcal{J}[u^*(t)] =
    \lim_{t \rightarrow t_k^+} \mathcal{J}[u^*(t)]
\end{equation}
at the boundaries. Recognizing that $\bar{n}(T)= \bar{N}$ leads to
the final Dolinar receiver error,
\begin{equation}
  P_\mathrm{D}(\eta) = \frac{1}{2}\left( 1-\sqrt{1-4 \xi_0 \xi_1
  c_0^{2 \eta}} \right)
\end{equation}
which is equal to the Helstrom bound for all values of the
detector efficiency, $0 < \eta \le 1$.

\section{Simulations} \label{Section::Simulations}

\begin{figure}[b]
\begin{center}
\includegraphics{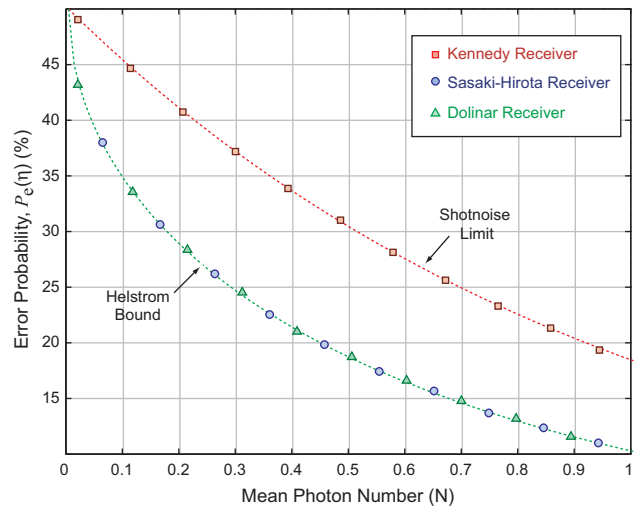}
\end{center}
\vspace{-6mm} \caption{Monte carlo simulation of the Kennedy,
Dolinar and Sasaki-Hirota receivers as a function of the signal
amplitude for perfect photon counting with
$\xi_0=\xi_1=\frac{1}{2}$. As expected, the Dolinar and
Sasaki-Hirota protocols both achieve the Helstrom bound while the
Kennedy receiver is approximately a factor of two worse.
 \label{Figure::AlphaScaling}}
\end{figure}

Monte Carlo simulations of the Kennedy, Sasaki-Hirota and Dolinar
receivers were performed to verify the above quantum efficiency
analysis and to analyze the effects of additional detector
imperfections.  Fig.\ \ref{Figure::AlphaScaling} shows benchmark
simulation results for perfect photodetection.  The three
receivers perform as expected in the small-amplitude regime; both
the Sasaki-Hirota and Dolinar protocols achieve the Helstrom bound
while the Kennedy receiver is approximately a factor of two worse,
at the shotnoise limit \footnote{In some contexts, Eq.\
(\ref{Equation::KennedyError}) is referred to as the
\textit{standard quantum limit} despite the fact that there is no
measurement backaction as $\hat{a}|\alpha\rangle = \alpha
|\alpha\rangle$.  We prefer the term \textit{shotnoise limit} in
order avoid such confusion.}. Statistics were accumulated for
10,000 Monte Carlo samples in which $\hat\rho_0$ and $\hat\rho_1$
were randomly selected with $\xi_0=\xi_1=1/2$.

\begin{figure}[t]
\begin{center}
\includegraphics{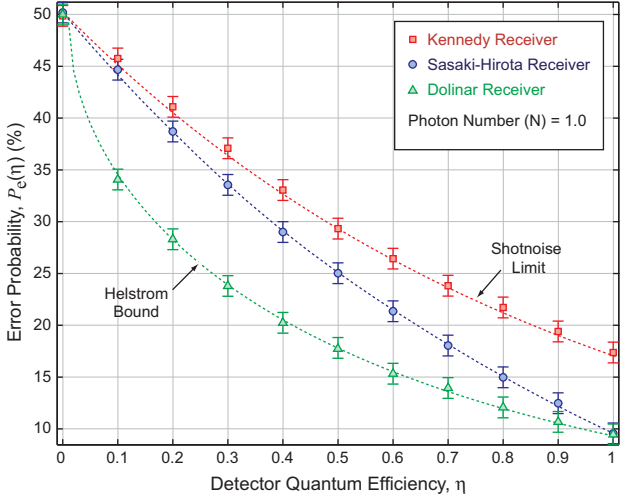}
\end{center}
\vspace{-6mm} \caption{Performance of the Kennedy, Sasaki-Hirota,
and Dolinar receivers as a function of the detector quantum
efficiency $\eta$.  The simulations were performed for
$\xi_0=\xi_1=\frac{1}{2}$ and data points reflect the result of
Monte Carlo simulations of the three receivers.  Dotted lines
correspond to the analytic results derived in Section
\ref{Section::CoherentStateCommunication} and illustrate that the
Dolinar receiver achieves the Helstrom bound even for sub-unity
quantum efficiency. \label{Figure::EtaScaling}}
\end{figure}

Detector imperfections, however, will degrade the performance of
each of the three receivers, and here we investigate the relative
degree of that degradation for conditions to be expected in
practice. The analysis is based on the observation that single
photon counting in optical communications is often implemented
with an avalanche photodiode (APD), as APDs generally provide the
highest detection efficiencies.  In the near-infrared, for
example, high-gain Silicon diodes provide a quantum efficiency of
$\eta\sim50$\%.  Additional APD non-idealities include: a dead
time following each detected photon during which the receiver is
unresponsive, dark counts in the absence of incoming photons due
to spontaneous breakdown events in the detector medium, a maximum
count rate above which the detector saturates (and can be
damaged), and occasional ghost clicks following a real photon
arrival--- a process referred to as ``after pulsing.'' For the
Dolinar receiver, which requires high-speed signal processing and
actuation in order to modulate the adaptive feedback field, delays
must also be considered.  That is, the optical modulators used to
adjust the phase and amplitude of the feedback signal as well as
the digital signal processing technology necessary
\cite{Stockton2002} for implementing all real-time computations
display finite bandwidths.

Fig.\ \ref{Figure::EtaScaling} compares the error probabilities of
the three receivers for sub-unity quantum efficiency but otherwise
ideal detection.  The mean photon number of the signal,
$\Psi_1(t)$, in this simulation is $\bar{N}=1$ with $\Psi_0(t)=0$
and $\xi_0=\xi_1=1/2$. Data points in the figure were generated by
accumulating statistics for 10,000 Monte Carlo simulations of the
three receivers, and the dotted lines correspond to the error
probabilities derived in Section
\ref{Section::CoherentStateCommunication}.  The simulations agree
well with the analytic expressions and it is evident that the
Dolinar receiver is capable of achieving the Helstrom bound for
$\eta<1$ while the Sasaki-Hirota receiver performance lies between
that of the Kennedy and Dolinar receivers.

\begin{figure}[t]
\begin{center}
\includegraphics{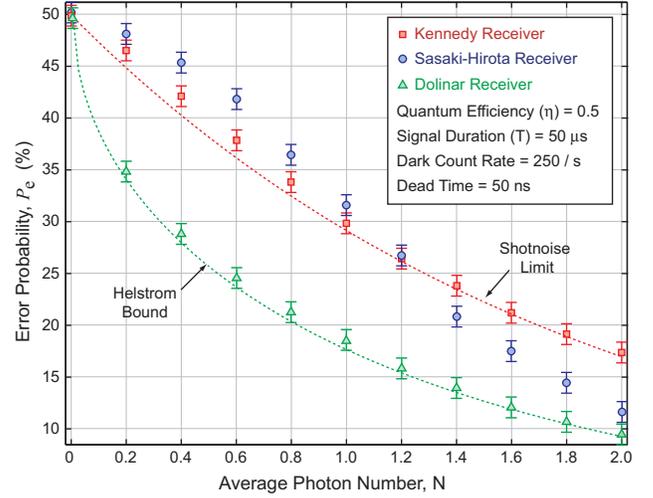}
\end{center}
\vspace{-6mm} \caption{Performance of the Kennedy, Sasaki-Hirota
and Dolinar receivers as a function of the mean number of photons
in the signal under realistic experimental conditions, including a
quantum efficiency of 50\%, a dark-count rate of 250 clicks/s, an
afterpulsing probability of 1\% and feedback delays of 100 ns.
\label{Figure::RealisticScaling}}
\end{figure}

Fig.\ \ref{Figure::RealisticScaling} compares the error
probabilities for the three receivers with the additional detector
and feedback non-idealities taken into account. Based on the
performance data of the Perkin-Elmer SPCM-AQR-13 Si APD single
photon counting module, we assumed a maximum count rate of $10^7$
photons/s, a detector dead-time of 50 ns, a dark count rate of 250
clicks/s and an after-pulsing probability of 1\%.  For the Dolinar
receiver, it was assumed that there was a 100 ns feedback delay
resulting from a combination of digital processing time and
amplitude/phase modulator bandwidth. The data points in Fig.\
\ref{Figure::RealisticScaling} correspond to the error
probabilities generated from 10,000 Monte Carlo simulations with
$\xi_0=\xi_1=1/2$.  The lower dotted line indicates the
appropriate Helstrom bound as a function of the mean photon
number, $\bar{N}$, for a detector with quantum efficiency,
$\eta=0.5$, and the upper curve indicates the analogous Kennedy
receiver error.  Evidentally, technical imperfections can have a
large negative effect on the performance of passive detection
protocols like the Kennedy and Sasaki-Hirota receivers while the
Dolinar receiver is more robust.

\begin{figure}
\includegraphics{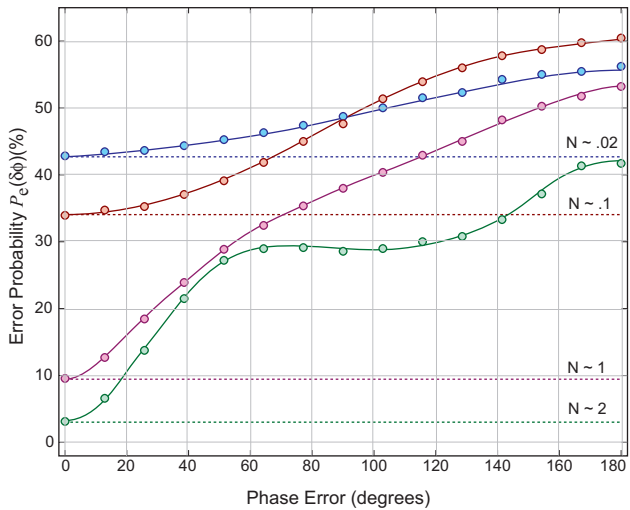}
\caption{Error probabilities for the Dolinar receiver as a
function of the phase error in the signal state corresponding to
logical 1 for different mean photon numbers.  Monte Carlo data
points were generated by accumulating statistics for 10,000 random
bits.  Solid lines are a fit to the data.
\label{Figure::PhaseError}}
\end{figure}

Unlike open-loop procedures, however, the feedback nature of the
Dolinar receiver additionally requires precise knowledge of the
incoming signal phase, $\varphi$, so that $U^*(t)$ can be properly
applied.  Fluctuations in the index of refraction of the
communication medium generally lead to some degree of phase noise
in the incoming signal, $\Psi(t)$. Adequately setting the phase of
$U(t)$ necessarily requires that some light from the channel be
used for phase-locking the local oscillator--- a task that reduces
the data transmission bandwidth. Therefore, operating a
communication system based on the Dolinar receiver at the highest
feasible rate requires that the number of photons diverted from
the data stream to track phase variations in the channel be
minimized. This optimization in turn requires knowledge of how
signal phase noise propagates into the receiver error probability.

Figure \ref{Figure::PhaseError} shows the error probability,
$P_\mathrm{D}(\delta\varphi)$ as a function of the phase
difference, $\delta\varphi$, between the incoming signal and the
local oscillator.  Data points correspond to results from 10,000
Monte Carlo simulations per photon number and phase angle, and the
solid curves reflect numerical fits to the Monte Carlo points. An
exact comparison between the open-loop and Dolinar receivers
requires information regarding the specific phase-error density
function for the actual communication channel being utilized.
However, we do note that at $\bar{N}=1$ photon, the phase of the
local oscillator could be as large as $\delta\varphi\sim 25^\circ$
before its error probability increased to that of the Kennedy
receiver.  Additionally, it appears that the slope of
$P_\mathrm{D}(\delta\varphi)$ is zero at $\delta\varphi=0$ which
implies that the Dolinar receiver conveniently displays minimal
sensitivity to small phase fluctuations in the channel.

\section{Discussion and Conclusions}

The Dolinar receiver was found to be robust to the types of
detector imperfections likely to exist in any real implementation
of a binary communication scheme based on optical coherent state
signaling and photon counting.  This robustness seemingly results
from the fact that the Dolinar receiver can correct itself after
events that cause an open-loop receiver to irreversibly
misdiagnose the transmitted state.  For example, imperfect
detection efficiency introduces a failure mode where the
probability, $p(H_0|\hat\rho_1)$, is increased above the value set
by quantum mechanical vacuum fluctuations. However, the optimal
structure of the Dolinar receiver feedback insures that it still
achieves the quantum mechanical minimum because it has control
over the counting rate.  That is, if the Dolinar receiver selects
the wrong hypothesis at some intermediate time, $t_k<T$, the
structure of the feedback insures that the receiver achieves the
highest allowable probability for invalidating that incorrect
decision during the remainder of the measurement, $t_k < t \le T$.

In the opposite situation, where dark counts or background light
produce detector clicks when there is no signal light in the
channel, open-loop receivers will decide in favor of $\hat\rho_1$
without any possibility for self-correction. This type of error
leads to an irreparable open-loop increase in $p(H_1|\hat\rho_0)$.
But, the Dolinar receiver has the potential to identify and fix
such a mistake since selecting the wrong hypothesis at
intermediate times increases the probability that a future click
will invalidate the incorrect decision.  When background light is
present, poor phase coherence between stray optical fields and the
signal provides no enhanced open-loop discrimination as there is
no local oscillator to establish a phase reference; a received
photon is a received photon (assuming that any spectral filtering
failed to prevent the light from hitting the detector).  The
Dolinar receiver is better immune to such an error since since
incoherent addition of the stray field to the local oscillator
will generally reduce the likelihood of a detector click, and even
if so, that click will be inconsistent with the anticipated
counting statistics.

Despite the previous belief that the Dolinar receiver is
experimentally impractical due to its need for real-time feedback,
we have shown that it is rather attractive for experimental
implementation. Particularly, quantum efficiency scales out of a
comparison between the Dolinar receiver error and the Helstrom
bound, while this is not the case for known unitary rotation
protocols. These results strongly suggest that real-time feedback,
previously cited as the Dolinar receiver's primary drawback, in
fact offers substantial robustness to many common imperfections
that would be present in a realistic experimental implementation.
Most importantly, simulations under these realistic conditions
suggest that the Dolinar receiver can out-perform the Kennedy
receiver with currently available experimental technology, making
it a viable option for small-amplitude, minimum-error optical
communication.

\begin{acknowledgements}
I would like to thank Hideo Mabuchi for countless insightful
comments and suggestions regarding this work and to acknowledge
helpful discussions with S. Dolinar and V. Vilnrotter. This work
was supported by the Caltech MURI Center for Quantum Networks
(DAAD-19-00-1-0374) and the NASA Jet Propulsion Laboratory. For
more information please visit http://minty.Caltech.EDU.
\end{acknowledgements}


\begin{thebibliography}{25}
\expandafter\ifx\csname
natexlab\endcsname\relax\def\natexlab#1{#1}\fi
\expandafter\ifx\csname bibnamefont\endcsname\relax
  \def\bibnamefont#1{#1}\fi
\expandafter\ifx\csname bibfnamefont\endcsname\relax
  \def\bibfnamefont#1{#1}\fi
\expandafter\ifx\csname citenamefont\endcsname\relax
  \def\citenamefont#1{#1}\fi
\expandafter\ifx\csname url\endcsname\relax
  \def\url#1{\texttt{#1}}\fi
\expandafter\ifx\csname
urlprefix\endcsname\relax\def\urlprefix{URL }\fi
\providecommand{\bibinfo}[2]{#2}
\providecommand{\eprint}[2][]{\url{#2}}

\bibitem[{\citenamefont{von Neumann}(1955)}]{vonNeumann1955}
\bibinfo{author}{\bibfnamefont{J.}~\bibnamefont{von Neumann}},
  \emph{\bibinfo{title}{Mathematical Foundations of Quantum Mechanics}}
  (\bibinfo{publisher}{Princeton University Press},
  \bibinfo{address}{Princeton}, \bibinfo{year}{1955}).

\bibitem[{\citenamefont{Holevo}(1973)}]{Holevo1973}
\bibinfo{author}{\bibfnamefont{A.~S.} \bibnamefont{Holevo}},
  \bibinfo{journal}{J. Multivar. Anal.} \textbf{\bibinfo{volume}{3}},
  \bibinfo{pages}{337} (\bibinfo{year}{1973}).

\bibitem[{\citenamefont{Helstrom}(1976)}]{Helstrom1976}
\bibinfo{author}{\bibfnamefont{C.~W.} \bibnamefont{Helstrom}},
  \emph{\bibinfo{title}{Quantum Detection and Estimation Theory}}, vol.
  \bibinfo{volume}{123} of \emph{\bibinfo{series}{Mathematics in Science and
  Egineering}} (\bibinfo{publisher}{Academic Press}, \bibinfo{address}{New
  York}, \bibinfo{year}{1976}).

\bibitem[{\citenamefont{Fuchs and Peres}(1996)}]{Fuchs1996}
\bibinfo{author}{\bibfnamefont{C.~A.} \bibnamefont{Fuchs}} \bibnamefont{and}
  \bibinfo{author}{\bibfnamefont{A.}~\bibnamefont{Peres}},
  \bibinfo{journal}{Phys. Rev. A} \textbf{\bibinfo{volume}{53}},
  \bibinfo{pages}{2038} (\bibinfo{year}{1996}).

\bibitem[{\citenamefont{Yuen and Shapiro}(1978)}]{Shapiro1978}
\bibinfo{author}{\bibfnamefont{H.~P.} \bibnamefont{Yuen}} \bibnamefont{and}
  \bibinfo{author}{\bibfnamefont{J.~H.} \bibnamefont{Shapiro}},
  \bibinfo{journal}{IEEE Trans. Inf. Theory} \textbf{\bibinfo{volume}{IT-24}},
  \bibinfo{pages}{657} (\bibinfo{year}{1978}).

\bibitem[{\citenamefont{Usuda and Hirota}(1995)}]{Hirota1995}
\bibinfo{author}{\bibfnamefont{T.~S.} \bibnamefont{Usuda}} \bibnamefont{and}
  \bibinfo{author}{\bibfnamefont{O.}~\bibnamefont{Hirota}},
  \emph{\bibinfo{title}{Quantum Communication and Measurement}}
  (\bibinfo{publisher}{Plenum}, \bibinfo{address}{New York},
  \bibinfo{year}{1995}).

\bibitem[{\citenamefont{Fuchs}(1997)}]{Fuchs1997}
\bibinfo{author}{\bibfnamefont{C.~A.} \bibnamefont{Fuchs}},
  \bibinfo{journal}{Phys. Rev. Lett.} \textbf{\bibinfo{volume}{79}},
  \bibinfo{pages}{1162} (\bibinfo{year}{1997}).

\bibitem[{\citenamefont{Peres and Wootters}(1991)}]{Peres1991}
\bibinfo{author}{\bibfnamefont{A.}~\bibnamefont{Peres}} \bibnamefont{and}
  \bibinfo{author}{\bibfnamefont{W.~K.} \bibnamefont{Wootters}},
  \bibinfo{journal}{Phys. Rev. Lett.} \textbf{\bibinfo{volume}{66}},
  \bibinfo{pages}{1119} (\bibinfo{year}{1991}).

\bibitem[{\citenamefont{Fuchs}(1996)}]{FuchsThesis}
\bibinfo{author}{\bibfnamefont{C.~A.} \bibnamefont{Fuchs}}, Ph.D. thesis,
  \bibinfo{school}{University of New Mexico} (\bibinfo{year}{1996}),
  \bibinfo{note}{quant-ph/9601020}.

\bibitem[{\citenamefont{Davies and Lewis}(1970)}]{Davies1970}
\bibinfo{author}{\bibfnamefont{E.~B.} \bibnamefont{Davies}} \bibnamefont{and}
  \bibinfo{author}{\bibfnamefont{J.~T.} \bibnamefont{Lewis}},
  \bibinfo{journal}{Comm. in Math. Phys.} \textbf{\bibinfo{volume}{17}},
  \bibinfo{pages}{239} (\bibinfo{year}{1970}).

\bibitem[{\citenamefont{Kraus}(1983)}]{Kraus1983}
\bibinfo{author}{\bibfnamefont{K.}~\bibnamefont{Kraus}},
  \emph{\bibinfo{title}{States, Efects, and Operations: Fundamental Notions of
  Quantum Theory}}, vol. \bibinfo{volume}{190} of
  \emph{\bibinfo{series}{Lecture Notes in Physics}}
  (\bibinfo{publisher}{Springer-Verlag}, \bibinfo{address}{Berlin},
  \bibinfo{year}{1983}).

\bibitem[{\citenamefont{Kennedy}(1972)}]{Kennedy1972}
\bibinfo{author}{\bibnamefont{Kennedy}}, \bibinfo{type}{Tech. Rep.}
  \bibinfo{number}{110}, \bibinfo{institution}{Research Laboratory of
  Electronics, MIT} (\bibinfo{year}{1972}).

\bibitem[{\citenamefont{Dolinar}(1973)}]{Dolinar1973}
\bibinfo{author}{\bibfnamefont{S.}~\bibnamefont{Dolinar}}, \bibinfo{type}{Tech.
  Rep.} \bibinfo{number}{111}, \bibinfo{institution}{Research Laboratory of
  Electronics, MIT} (\bibinfo{year}{1973}).

\bibitem[{\citenamefont{Belavkin et~al.}(1995)\citenamefont{Belavkin, Hirota,
  and Hudson}}]{Belavkin1995}
\bibinfo{author}{\bibfnamefont{V.~P.} \bibnamefont{Belavkin}},
  \bibinfo{author}{\bibfnamefont{O.}~\bibnamefont{Hirota}}, \bibnamefont{and}
  \bibinfo{author}{\bibfnamefont{L.}~\bibnamefont{Hudson}},
  \emph{\bibinfo{title}{Quantum Communication and Measurement}}
  (\bibinfo{publisher}{Plenum}, \bibinfo{address}{New York},
  \bibinfo{year}{1995}).

\bibitem[{\citenamefont{Monmose et~al.}(1996)\citenamefont{Monmose, Osaki, Ban,
  Sasaki, and Hirota}}]{Hirota1996a}
\bibinfo{author}{\bibfnamefont{R.}~\bibnamefont{Monmose}},
  \bibinfo{author}{\bibfnamefont{M.}~\bibnamefont{Osaki}},
  \bibinfo{author}{\bibfnamefont{M.}~\bibnamefont{Ban}},
  \bibinfo{author}{\bibfnamefont{M.}~\bibnamefont{Sasaki}}, \bibnamefont{and}
  \bibinfo{author}{\bibfnamefont{O.}~\bibnamefont{Hirota}}, in
  \emph{\bibinfo{booktitle}{NASA Proc. Ser.}} (\bibinfo{year}{1996}).

\bibitem[{\citenamefont{Sasaki and Hirota}(1996)}]{Hirota1996b}
\bibinfo{author}{\bibfnamefont{M.}~\bibnamefont{Sasaki}} \bibnamefont{and}
  \bibinfo{author}{\bibfnamefont{O.}~\bibnamefont{Hirota}},
  \bibinfo{journal}{Phys. Rev. A} \textbf{\bibinfo{volume}{54}},
  \bibinfo{pages}{2728} (\bibinfo{year}{1996}).

\bibitem[{\citenamefont{Armen et~al.}(2002)\citenamefont{Armen, Au, Stockton,
  Doherty, and Mabuchi}}]{Armen2002}
\bibinfo{author}{\bibfnamefont{M.~A.} \bibnamefont{Armen}},
  \bibinfo{author}{\bibfnamefont{J.~K.} \bibnamefont{Au}},
  \bibinfo{author}{\bibfnamefont{J.~K.} \bibnamefont{Stockton}},
  \bibinfo{author}{\bibfnamefont{A.~C.} \bibnamefont{Doherty}},
  \bibnamefont{and} \bibinfo{author}{\bibfnamefont{H.}~\bibnamefont{Mabuchi}},
  \bibinfo{journal}{Phys. Rev. Lett.} \textbf{\bibinfo{volume}{89}},
  \bibinfo{pages}{133602} (\bibinfo{year}{2002}).

\bibitem[{\citenamefont{Geremia et~al.}(2004)\citenamefont{Geremia, Stockton,
  and Mabuchi}}]{Geremia2004a}
\bibinfo{author}{\bibfnamefont{J.}~\bibnamefont{Geremia}},
  \bibinfo{author}{\bibfnamefont{J.~K.} \bibnamefont{Stockton}},
  \bibnamefont{and} \bibinfo{author}{\bibfnamefont{H.}~\bibnamefont{Mabuchi}},
  \bibinfo{journal}{Science} \textbf{\bibinfo{volume}{304}},
  \bibinfo{pages}{270} (\bibinfo{year}{2004}).

\bibitem[{\citenamefont{JM~GEremia and Mabuchi}(2004)}]{Geremia2004b}
\bibinfo{author}{\bibfnamefont{J.~K.~S.} \bibnamefont{JM~GEremia}}
  \bibnamefont{and} \bibinfo{author}{\bibfnamefont{H.}~\bibnamefont{Mabuchi}}
  (\bibinfo{year}{2004}), \bibinfo{note}{quant-ph/0401107}.

\bibitem[{\citenamefont{Peres}(1990)}]{Peres1990}
\bibinfo{author}{\bibfnamefont{A.}~\bibnamefont{Peres}},
  \bibinfo{journal}{Found. Physics} \textbf{\bibinfo{volume}{20}},
  \bibinfo{pages}{1441} (\bibinfo{year}{1990}).

\bibitem[{\citenamefont{Helstrom}(1967)}]{Helstrom1967}
\bibinfo{author}{\bibfnamefont{C.~W.} \bibnamefont{Helstrom}},
  \bibinfo{journal}{Information and Control} \textbf{\bibinfo{volume}{10}},
  \bibinfo{pages}{254} (\bibinfo{year}{1967}).

\bibitem[{\citenamefont{Glauber}(1963)}]{Glauber1963}
\bibinfo{author}{\bibfnamefont{R.~J.} \bibnamefont{Glauber}},
  \bibinfo{journal}{Phys. Rev.} \textbf{\bibinfo{volume}{131}},
  \bibinfo{pages}{2766} (\bibinfo{year}{1963}).

\bibitem[{\citenamefont{Dolinar}(1976)}]{DolinarThesis}
\bibinfo{author}{\bibfnamefont{S.}~\bibnamefont{Dolinar}}, Ph.D. thesis,
  \bibinfo{school}{Massachussets Institute of Technology}
  (\bibinfo{year}{1976}).

\bibitem[{\citenamefont{Bertsekas}(2000)}]{Bertsekas2000}
\bibinfo{author}{\bibfnamefont{D.~P.} \bibnamefont{Bertsekas}},
  \emph{\bibinfo{title}{Dynamic Programming and Optimal Control}},
  vol.~\bibinfo{volume}{1} (\bibinfo{publisher}{Athena Scientific},
  \bibinfo{address}{Belmont, Massachusetts}, \bibinfo{year}{2000}),
  \bibinfo{edition}{2nd} ed.

\bibitem[{\citenamefont{Stockton et~al.}(2002)\citenamefont{Stockton, Armen,
  and Mabuchi}}]{Stockton2002}
\bibinfo{author}{\bibfnamefont{J.}~\bibnamefont{Stockton}},
  \bibinfo{author}{\bibfnamefont{M.}~\bibnamefont{Armen}}, \bibnamefont{and}
  \bibinfo{author}{\bibfnamefont{H.}~\bibnamefont{Mabuchi}},
  \bibinfo{journal}{J. Opt. Soc. Am. B} \textbf{\bibinfo{volume}{19}},
  \bibinfo{pages}{3019} (\bibinfo{year}{2002}).

\end{thebibliography}

\end{document}